%% file: wavelet.tex
\documentclass[10pt,journal]{IEEEtran}

\usepackage[T1]{fontenc}
\usepackage{amsmath}
\usepackage{amssymb}
\usepackage[mathscr]{eucal}
\usepackage{mathtools}
\usepackage{bm}
\usepackage{pdfsync}  % this messes up and is messed up by lineno
\usepackage[hidelinks]{hyperref}
\usepackage{subcaption}

\usepackage{xcolor}

\DeclareMathAlphabet{\mathpzc}{OT1}{pzc}{m}{it}

\newcommand{\etal}{\textit{et al.\ }}

\begin{document}

%\iflinenumbers \linenumbers \fi

\title{Bayesian Functional Mixed-Effects Model with Gaussian Process Responses for Wavelet Spectra of Spatiotemporal Colonic Manometry Signals}

\author{Lukasz Wiklendt, Marcello Costa, Simon Brookes, Phil G. Dinning \thanks{L. Wiklendt, M. Costa, and S. Brookes are with the College of Medicine and Public Health, Flinders University, Australia.} \thanks{PG. Dinning is with the Department of Surgery, Flinders Medical Centre, Australia.} \thanks{Received grant supported by the National Health and Medical Research Council of Australia (ID: 1064835, GNT1162223) to conduct the work in this paper.}\thanks{Ethics for the studies was obtained from the South Adelaide Health Service/Flinders University Human Research Ethics Committee (409.10; March 2011).}}

\maketitle

%\begin{abstract}
%We present a technique for identification and statistical analysis of quasiperiodic spatiotemporal pressure signals recorded from multiple closely spaced sensors in the human colon. Identification is achieved by computing the continuous wavelet transform and cross-wavelet transform of these recorded signals. Statistical analysis is achieved by modelling the resulting time-averaged amplitudes or coherences in the frequency and frequency-phase domains as Gaussian processes over a regular grid, under the influence of categorical and numerical predictors that are specified by the experimental design as a mixed-effects model. Parameters of the model are inferred with Hamiltonian Monte Carlo. We present an application of this method to determine statistical differences in pressure signals between different colonic regions and the change in these signals in response to a meal.
%\end{abstract}

\begin{abstract}
\emph{Objective:} We present a technique for identification and statistical analysis of quasiperiodic spatiotemporal pressure signals recorded from multiple closely spaced sensors in the human colon. \emph{Methods:} Identification is achieved by computing the continuous wavelet transform and cross-wavelet transform of these recorded signals. Statistical analysis is achieved by modelling the resulting time-averaged amplitudes or coherences in the frequency and frequency-phase domains as Gaussian processes over a regular grid, under the influence of categorical and numerical predictors that are specified by the experimental design as a mixed-effects model. Parameters of the model are inferred with Hamiltonian Monte Carlo. \emph{Results and Conclusion:} We present an application of this method to colonic manometry data in healthy controls, to determine statistical differences in the spectra of pressure signals between different colonic regions and in response to a meal intervention. We are able to successfully identify and contrast features in the spectra of pressure signals between various predictors. \emph{Significance:} This novel method provides fast analysis of manometric signals at levels of detail orders of magnitude beyond what was previously available. The proposed tractable mixed-effects model is broadly applicable to experimental designs with functional responses.
\end{abstract}


\section{Introduction}

Gut motility describes the coordinated pattern of contractions and relaxations of the longitudinal and circular smooth muscle layers of the gut wall.  This motility plays a central role in normal digestive health, mixing and propelling gut content in a controlled fashion. Motility disorders in children and adults cause significant personal, societal and financial burdens, costing the healthcare systems many billions of dollars per year \cite{peery2018burden}. In order to improve treatment options, we need to gain insight into the changes in gut dysmotility associated with a disorder. For example, in the human colon, large propulsive propagating contractions have been shown to be absent or diminished in patients with severe constipation \cite{bassotti1988colonic}. In all regions of the gut, muscle contractions can be detected using intraluminal manometry. This procedure involves placement, through the nose or anus, of a flexible, small diameter catheter, containing multiple sensors. In recent years the number of sensors on these catheters has increased dramatically. For colonic manometry recordings, these high-resolution manometry catheters have allowed for a far greater insight into the nature of normal \cite{dinning2014quantification} and abnormal \cite{dinning2015colonic} contractile motor patterns. Analysis of these data is difficult. Manometric recordings consist of peaks of pressure which reflect contractile activity of the intestinal muscle. While, visual analysis can easily identify patterns of motor activity, the process is very time consuming, unstandardised, subject to personal bias, and in some instances the sheer volume of pressure peaks, occurring at multiple frequencies across many sensors can be overwhelming.

% Phil's version, abridged
%Gut motility describes the coordinated pattern of contractions and relaxations of the smooth muscle layers of the gut wall, which can be recorded using intraluminal manometry. This procedure involves placement of a flexible, small diameter catheter, containing multiple sensors into the colon \cite{dinning2018new}. Analysis of these data is difficult, in many instances, patterns of contractile activity are identified manually by visual analysis. With the increase in data collated with high-resolution manometry, the ability to visually identify all individual patterns becomes increasingly difficult (Fig. \ref{fig:wavelet}a). This paper introduces an automated analysis and statistical technique for processing manometry data.

\begin{figure*}
	\centering
	\includegraphics[width=\textwidth]{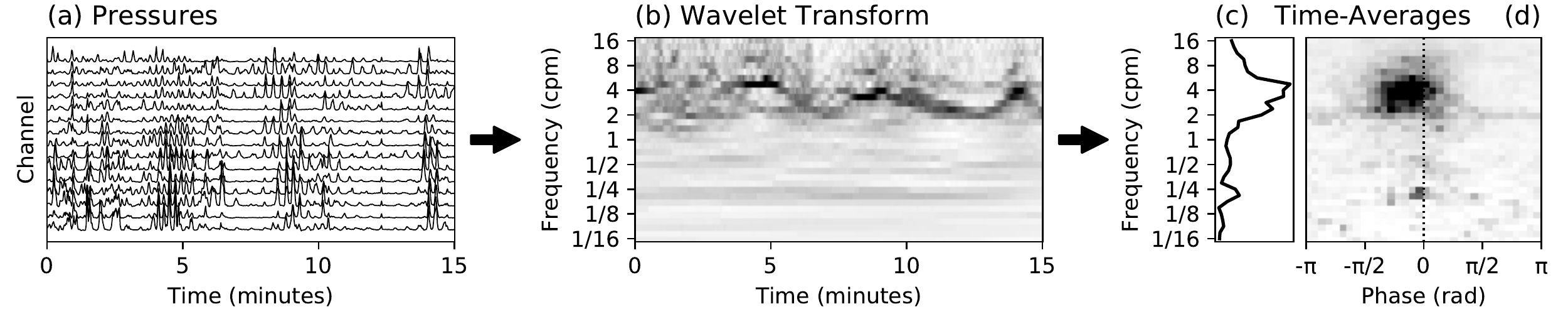}
	%\iflinenumbers \internallinenumbers \fi
	\caption{Wavelet computation process. (a) Pressures were recorded by a high-resolution manometry catheter from the colon of a healthy adult. The sensors on the catheter are spaced evenly at 10mm intervals. This particular example shows data recorded from 15 sensors over a 15-minute period, where channels are arranged such that the top channel is most oral and bottom channel most aboral. (b) The wavelet transform (shown as an average over channels) and cross-wavelet transform (not shown) are computed, with 33 logarithmically-spaced frequencies shown from 1/16 to 16 cycles-per-minute (cpm). (c) The time-average of the wavelet transform is computed and (d) the time-average of the cross-wavelet transform stratified by phase is computed, shown with a vertical dotted line indicating 0-phase corresponding to synchronous activity. A typical 2-6 cpm frequency is observed in the upper half of (d) as a large dark feature mostly on the left side of the 0-phase line, indicating a predominantly retrograde (oral) direction of propagation. The 2-6 cpm activity occurs in bouts of approximately one every 4-5 minutes, which can be observed as a smaller feature in the lower half of (d) at the 0-phase line.}
	\label{fig:wavelet}
\end{figure*}

An automated approach to overcome these issues is needed. In this paper, a two-stage process to achieve this is presented. The first stage involves computing a frequency spectrum of the spatiotemporal pressure recordings with the wavelet transform. The second stage compares the spectra via a functional mixed-effects model to reveal where differences over frequencies may lie in response to various known predictors (e.g. response to a meal, age, colonic region). Although we focus on colonic manometry, the two stages are applicable to other kinds of recorded phenomena of a similar quasiperiodic nature. The second stage is also widely applicable to data where each observation can be represented as a function over a shared gridded domain.

The continuous wavelet transform \cite{mallat2008wavelet} converts a signal from the time domain to the time-frequency\footnote{The transformation is actually to the time-scale domain, where an appropriate conversion from scale to frequency can be chosen.} domain, such that for any point in time we get a distribution of frequencies (\emph{wavelet coefficients}) occurring in the signal at that time (Fig \ref{fig:wavelet}b).

%There is also a discrete wavelet transform counterpart, but it is less suitable for interpretable signal analysis \cite{lau1995climate}, and excludes important features such as phase measurement that is available in the continuous version.

The application of the wavelet transform for the analysis of acquired data have been used in many disciplines, from the recording of pressure signals in the human rectum \cite{jiang2012optimal} to analysis of black hole gravitational waves \cite{abbott2016observation}. In particular, the wavelet transform has many generally applicable expositions can be found in the geophysical sciences \cite{liu1994wavelet, lau1995climate, torrence1998practical, torrence1999interdecadal, grinsted2004application}. In this paper, we focus on the time-averaged wavelet coefficients, including 2D histograms over frequencies and phase-differences between pairs of signals to get an estimate of propagation of quasiperiodic pressure signals. This is similar to the work done on the solar atmosphere using the FFT \cite{lites1979vertical} and wavelets \cite{andic2008propagation}.

For statistical analysis, we use function-on-scalar rather than the more conventional scalar-on-scalar regression. Scalar-on-scalar regression covers the commonly encountered regression such as the two-sample t-test, ANOVA, or generalised linear model. Observations from this framework consist of a scalar (single number) response, and a set of scalar predictors encoding the experimental conditions (e.g. response to stimulus) and individual characteristics of each subject being recorded (e.g. age or sex). Function-on-scalar regression extends this so that responses are not scalars but functions, such as frequency spectra. See \cite{morris2015functional} for a review on functional regression.

There are many previous publications on functional data analysis, with one of the first text books on the topic published in 1997 \cite{ramsay1997functional}. Functions in mixed-effects models are often represented by sums of basis functions, such as splines \cite{guo2002functional, shi2012mixed, yang2017efficient}, sines and cosines \cite{diggle1997spectral, bose2018toward}, or discrete wavelet bases \cite{morris2003wavelet, morris2006wavelet, morris2008bayesian, morris2011automated, martinez2013study}. Non-parametric forms such as Gaussian processes (GP) are more expressive but computationally expensive. GPs have seen a slower adoption in functional mixed-effects models, with simpler designs such as a single function \cite{goldberg1998regression}, fixed-effects models \cite{shi2005hierarchical}, or simple hierarchical models \cite{yang2016smoothing}.

%Sometimes, authors use basis-function approximations of GPs for computational tractability. Although technically still GPs, such approximations come at the cost of limiting the functional form to the particular kernel induced by the approximation. A general GP model allows one to specify the kernel function itself, rather than needing to search for basis functions to reverse engineer a desired kernel.

% Unfortunately, many authors claim their mixed-effects models are GP-based, but then resort to basis-function approximations for computational tractability. Although technically still GPs, such approximations come at the cost of limiting the functional form to the particular kernel induced by the approximation. A general GP model allows one to specify the kernel function itself, rather than needing to search for basis functions to reverse engineer a desired kernel.

When considering a single function model, authors have explored more complex GP structures such as heteroscedasticity and other non-stationarities. Goldberg \etal \cite{goldberg1998regression} and Titsias and L'azaro-Gredilla \cite{titsias2011variational} presented a GP model with a GP-based input-dependent noise variance. Tolvanen \etal \cite{tolvanen2014expectation} presented a GP with a GP-based input-dependent signal and noise variance. They noted that input-dependent lengthscale and signal variance are underidentifiable, and opted for an input-dependent signal variance. Despite this underidentifiability, Heinonen \etal \cite{heinonen2016non} successfully fit a GP with a squared-exponential kernel function with GP-based input-dependent signal variance, noise variance, and lengthscale.

\subsection*{Novelty}

Here we present a heteroscedastic Gaussian Process mixed-effect model with Gaussian Process residuals over a common grid. A substantial performance improvement is achieved on a 2D grid by exploiting separable kernels via Kronecker factorisation, inspired by \cite{flaxman2015fast}. The model is conveniently written in the Stan \cite{carpenter2017stan} probabilistic programming language.

We also present a novel amplitude-weighted coherence summary of the cross-wavelet transform useful for focusing on evident activity by effectively down-weighting periods of quiescence, when little to no contractile activity is present.

\subsection*{Layout}

This paper is organised as follows. The wavelet transform is summarised in section \ref{sec:wt}, including the calculation of global wavelet power spectrum used for inference. Section \ref{sec:xwt} describes how the wavelet spectra from two signals can be combined to calculate the wavelet cross-spectrum and the associated coherence, also outlining the calculation of histograms over phase-differences from the power cross-spectra and coherence. Section \ref{sec:stats} presents a method for statistically modelling the wavelet results as a heteroscedastic functional mixed-effects model on the basis of Gaussian processes, with inference conducted by Hamiltonian Monte Carlo (HMC) using the Stan software ecosystem \cite{carpenter2017stan}. We apply our method to signals recorded with colonic manometry in 11 healthy volunteers, presenting the results in section \ref{sec:application}, and conclude with a discussion in section \ref{sec:discussion}.

\section{Wavelet Transform} \label{sec:wt}

The continuous wavelet transform \cite{mallat2008wavelet, torrence1998practical} is a useful tool for analysing non-stationary quasiperiodic signals. It decomposes a time domain signal $x(t) \in \mathbb{R}$ into the time-scale domain $w(t, s) \in \mathbb{C}$ with
\begin{eqnarray}
w(t,s) &=& \int_{-\infty}^\infty x(\tau) \frac{1}{\sqrt{s}}\psi^*\left(\frac{\tau - t}{s}\right) d\tau \label{eqn:cwt}
\end{eqnarray}
where $\psi(t) \in \mathbb{C}$ is an admissable \emph{wavelet} function, and the $*$ superscript represents the complex conjugate. An admissible wavelet function is one which has zero mean and its Fourier transform is continuously differentiable \cite{farge1992wavelet}, with an extra desirable property that it be localised in both time and frequency. Intuitively, $w(t,s)$ measures the variation of $x(t)$ within a neighbourhood at $t$ of size proportional to $s$.

In practice, we choose $s$ from finite set of logarithmically-spaced scales $S = \{s_1, \dotsc, s_L\}$, specify the wavelet basis function in the frequency domain, and perform the convolution in (\ref{eqn:cwt}) via FFT utilising the convolution theorem with
\begin{eqnarray}
	w(t, s) &=& \mathcal{F}^{-1}\left[X(\omega)\sqrt{s} \Psi^*(s \omega)\right] (t)
\end{eqnarray}
where $s \in S$, $\Psi = \mathcal{F}[\psi]$ is frequency-domain wavelet function, $X = \mathcal{F}[x]$ is the frequency-domain signal, $\mathcal{F}$ and $\mathcal{F}^{-1}$ are the Fourier and inverse Fourier transforms, and $\omega$ represents the frequency-domain locations in radians per second.

% If a recorded signal is not in cyclical time, that is, if the first and one-past-last time samples are not identified, then edge effects of the recording can creep in. In such cases the ends of $w$ should be set to an amplitude of $0$ at a distance proportional to $s$ from each edge, since those samples are under the influence of edge effects. This region is known as the \emph{cone-of-influence}. See \cite{torrence1998practical} for a tutorial on the wavelet transform that includes sufficient practical details for implementation.

To map from scales (seconds) to frequencies (Hz) we use ``Synchrosqueezing'' \cite{daubechies2011synchrosqueezed}. Synchrosqueezing redistributes the wavelet coefficients based on the first time-derivative of the phase (also known as ``instantaneous frequency''). For a given set of $K$ equally-and-logarithmically spaced positive frequency bins with ordered edges $\{\mathpzc{f}_1, \dotsc, \mathpzc{f}_{K+1}\}$, where $\mathpzc{F} = \{[\mathpzc{f}_1, \mathpzc{f}_2), [\mathpzc{f}_2, \mathpzc{f}_3), \dotsc, [\mathpzc{f}_K, \mathpzc{f}_{K+1})\}$ is the set of half-open intervals characterising each bin's domain, synchrosqueezing can be described as
\begin{eqnarray}
	v(t,f) &=& \sum_{s\in S} \frac{w(t,s)}{\sqrt{s}} \mathbf{1}_{\mathrm{bin}_\mathpzc{F}(f)}\left(\frac{1}{2\pi}\frac{\partial \phi(t,s)}{\partial t}\right) \label{eqn:syncsqz} \\
	\mathrm{bin}_\mathpzc{A}(x) &\coloneqq& \{ a \in A\ |\ (\exists A \in \mathpzc{A})[x \in A] \} \\
	\mathbf{1}_\Omega(x) &\coloneqq&
		\begin{cases}
			1 & \text{if } x \in \Omega \\
			0 & \text{if } x \notin \Omega
		\end{cases}
\end{eqnarray}
where $\phi(t,s) = \mathrm{unwrap}(\angle w(t,s))$ represents the time-differentiable ``unwrapped-in-time'' phase in radians with the complex argument (or angle) denoted by the parentheses-less function $\angle$ with domain-codomain $\angle: \mathbb{C} \rightarrow [-\pi, \pi)$. The function $\mathrm{bin}_\mathpzc{F}(f): \mathbb{R} \rightarrow \mathpzc{F}$ returns the interval from $\mathpzc{F}$ that contains $f$, and $\mathbf{1}$ is the indicator function. The equally-and-logarithmically spaced condition on the bins can be satisfied with a constant $\Delta_\mathpzc{F}$, where $\log \Delta_\mathpzc{F} = \log \mathpzc{f}_{k+1} - \log \mathpzc{f}_k$ for all $[\mathpzc{f}_k, \mathpzc{f}_{k+1}) \in \mathpzc{F}$. An example is shown in Fig. \ref{fig:wavelet}b.

Let $\mathpzc{F}_k = [\mathpzc{f}_k, \mathpzc{f}_{k+1})$ be an element of $\mathpzc{F}$. Since $\forall \mathpzc{F}_k \in \mathpzc{F}$ and $\forall f, f' \in \mathpzc{F}_k$ the equality $v(t, f) = v(t, f')$ clearly holds, then we only need to consider a single arbitrary $f \in \mathpzc{F}_k$ to represent bin $k$. We will use the logarithmic-centre of each bin, and define the set of bin centres as
\begin{eqnarray}
	F &=& \{f_1, \dotsc, f_k, \dotsc, f_K\} \label{eqn:F}\\
	f_k &=& e^{\frac{1}{2}(\log \mathpzc{f}_{k} + \log \mathpzc{f}_{k+1})}
\end{eqnarray}

Switching to discrete-time representation with samples recorded at times $T = \{t_1, \dotsc, t_N\}$ we can view the wavelet spectrum as $v(t, f): T \times F \rightarrow \mathbb{C}$. The time-average of the squared amplitudes produces the global wavelet power spectrum
\begin{eqnarray}
	\hat{v}(f)^2 &=& \frac{1}{N} \sum_{t\in T} \left|v(t, f)\right|^2 \label{eqn:gws}
\end{eqnarray}

An example is shown in Fig. \ref{fig:wavelet}c.

\section{Cross-Wavelet Transform} \label{sec:xwt}

The cross-wavelet transform combines two wavelet spectra with the complex-conjugated product
\begin{eqnarray}
	v_{ab}(t,f) &=& v_a(t,f) v_b^*(t,f) \label{eqn:xwt}
\end{eqnarray}
where $v_a$ and $v_b$ are the synchrosqueezed wavelet transforms of the two signals labelled $a$ and $b$. The combined subscript $v_{ab}$ denotes the cross-wavelet transform between the two signals.

A global wavelet power cross-spectrum could be computed in the same way for $v_{ab}$ as shown for $v$ in equation (\ref{eqn:gws}). However, this discards the useful phase information contained in $v_{ab}$. The effect of the complex-conjugated product is that the resulting phase represents the difference in phase between the two signals. For each frequency, computing a squared-amplitude-weighted histogram of the phase-differences yields a 2D histogram in the frequency-phase domain, analogous to the global wavelet power spectrum but stratified by phase-differences.

%Since phase-differences are actually phases, in the rest of this section we will refer to them simply as ``phases'', keeping in mind that they represent the phase-difference between two signals, rather than the phase of one or the other.

Given a set of $M$ equally-spaced bins with ordered edges $\{\mathpzc{h}_1, \dotsc, \mathpzc{h}_{M+1}\}$ representing phase-differences, where $\mathpzc{H} = \{[\mathpzc{h}_1, \mathpzc{h}_2), [\mathpzc{h}_2, \mathpzc{h}_3), \dotsc, [\mathpzc{h}_{M}, \mathpzc{h}_{M+1})\}$ is the set of half-open intervals characterising each bin's domain, we define the 2D histogram of frequencies and phase-differences by
\begin{eqnarray}
	\hat{v}_{ab}(f, \varphi) &=& \frac{\sum_{t \in T_{v_{ab}}(f, \varphi)} |v_{ab}(t,f)|}{|T_{v_{ab}}(f, \varphi)|} \label{eqn:v_fp} \\
	T_z(f, \varphi) &\coloneqq& \{ t \in T\ |\ \angle z(t,f) \in \mathrm{bin}_\mathpzc{H}(\varphi) \} \label{eqn:T_Phi}
\end{eqnarray}
where $\mathrm{bin}_\mathpzc{H}(\varphi): \mathbb{R} \rightarrow \mathpzc{H}$ returns the interval from $\mathpzc{H}$ that contains $\varphi$, and $T_z(f, \varphi)$ is the set of all the time samples such that $\angle z(t, f)$ is in the bin containing $\varphi$. To ensure all phases are covered $\mathpzc{h}_1 = -\pi$ and $\mathpzc{h}_{M+1} = \pi$. Let $\mathpzc{H}_m = [\mathpzc{h}_{m}, \mathpzc{h}_{m+1})$ be an element of $\mathpzc{H}$. Since $\forall \mathpzc{H}_m \in \mathpzc{H}$ and $\forall \varphi, \varphi' \in \mathpzc{H}_m$ the equality $\hat{v}_{ab}(f, \varphi) = \hat{v}_{ab}(f, \varphi')$ clearly holds, then we only need to consider a single arbitrary $\varphi \in \mathpzc{H}_m$ to represent bin $m$. We will use the centre of each bin, defining the set of bin centres as
\begin{eqnarray}
	H &=& \{\varphi_1, \dotsc, \varphi_m, \dotsc, \varphi_M\} \label{eqn:H} \\
	\varphi_m &=& \frac{1}{2}(\mathpzc{h}_{m} + \mathpzc{h}_{m+1})
\end{eqnarray}

If pairs of sensors are spaced sufficiently close together in the environment being recorded, then the cross-wavelet transform between sensors in such a pair allows us to measure propagating quasiperiodic activity. The sign of the phase-difference determines the direction of propagation. An example is shown in Fig. \ref{fig:wavelet}d, where the dark feature in the upper half of the panel and slightly to the left of the dotted centre line indicates a strong tendency of the pressure waves in Fig. \ref{fig:wavelet}a to move bottom-to-top, i.e. in an oral direction. The value of the phase-difference $\varphi$ (rad) at the frequency of interest $f$ (Hz) and the separation between the pair of sensors $d$ (cm) can be used to determine the apparent velocity of propagation $u$ (cm/s) with the simple formula
\begin{eqnarray}
u &=& d\ 2\pi f /\varphi \label{eqn:velocity}
\end{eqnarray}
For quasiperiodic pressure signals in these data, a more appropriate measure of propagation may be ``pace'' which is the inverse velocity $u^{-1}$ (s/cm), where synchronous events (or phase-locking) between the two signals have a more robust-for-modelling pace of 0, rather than a velocity at $\pm \infty$.

\subsection*{Coherence}

If the distance between pairs of sensors is too large in relation to the propagation speed and frequency of the phenomena being recorded, then aliasing occurs whenever events are out of step with each other by more than half a cycle. This causes the phase-difference to misrepresent the true direction of propagation and its apparent velocity. In this case coherence can be used as a measure of coordination between pairs of sensors, even if we are forfeiting a measure of velocity. Coherence is a value between 0 and 1 that measures the coefficient-of-determination between two time series as a function of time and frequency \cite{liu1994wavelet}.

To calculate the coherence, the method in \cite{torrence1999interdecadal} uses the power cross-spectrum $|w_a w^*_b|^2$ divided by each of the individual power spectra $|w_a|^2$ and $|w_b|^2$. However, this is always identical to 1, and so \cite{torrence1999interdecadal} suggest smoothing in time and scale the cross-spectrum and the individual power spectra before calculating the coherence. However, since we are using the synchronsqueezed coefficients $v_a$ and $v_b$, then we smooth only in time and not scale, because the frequency-based redistribution of coefficients makes smoothing in scale no longer relevant. Here, we measure coherence as
\begin{eqnarray}
	R_{ab}^2(t,f) &=& \frac{|\left< v_{ab}(t,f) \right>_f|^2}{\left<| v_a(t,f) |^2\right>_f \left<| v_b(t,f) |^2\right>_f} \label{eqn:R2tf}
\end{eqnarray}
where the angle brackets $\left<\ \right>_f$ denote smoothing-in-time, allowing for an $f$-dependent smoothing width. The smoothing width is somewhat arbitrary \cite{torrence1999interdecadal}, but one can still tailor it to the wavelet function and perhaps the phenomena being recorded. For the examples presented in this paper we use Gaussian smoothing with a width parameter of $\sigma = 4 f^{-1}$ since we're using a Morlet wavelet function with an intrinsic frequency of $\omega_0 = 6$ radians, which corresponds to approximately 4 oscillations occurring within about 95\% of the wavelet's Gaussian envelope.

It is clear that (\ref{eqn:R2tf}) is no greater than 1 due to Jensen's inequality, and no less than 0 due to the squared modulus. A shorter smoothing width will tend to produce higher coherence since the region over which the signal influences the coherence measurement is shorter. In the limit of no smoothing, substituting (\ref{eqn:xwt}) into (\ref{eqn:R2tf}) yields 1. In the limit of total smoothing or averaging over the entire time range, and assuming that $v_{ab}$ is zero-mean random processes, then (\ref{eqn:R2tf}) will tend to 0.

To obtain a global measure of coherence, we use the logit-transformed squared-amplitude-weighted average over time
\begin{eqnarray}
	\hat{R}_{ab}^2(f) &=& \mu_T^\mathrm{logit}\left[ R_{ab}^2(\cdot, f), |v_{ab}(\cdot, f)|^2 \right] \label{eqn:R2f} \\
	\mu_\Omega^g[\alpha, p] &\coloneqq& g^{-1}\left( \frac{\sum_{x \in \Omega} p(x) g(\alpha(x))}{\sum_{x \in \Omega} p(x)} \right) \label{eqn:wavg}
\end{eqnarray}
where $\mu$ calculates the $g$-transformed average of $\alpha: \Omega \rightarrow \mathbb{R}$ weighted by $p: \Omega \rightarrow \mathbb{R}$, over the domain $\Omega$. A centre dot in a function denotes the variable of a curried function, that is, for some function $q$ we have $q(\cdot,y)(x) = q(x,y)$. Utilising $\mu$, equation (\ref{eqn:v_fp}) could be written
\begin{eqnarray}
\hat{v}_{ab}(f, \varphi) &=& \mu_{T_{v_{ab}}(f, \varphi)}^\mathrm{id}\left[|v_{ab}(\cdot, f)|, 1 \right] \nonumber
\end{eqnarray}
where $\mathrm{id}$ is the identity function.

Similarly to $v_{ab}(t,f)$ in equation (\ref{eqn:v_fp}) we can calculate the histogram of $R_{ab}^2(t,f)$ over phases, after a logit transformation and weighting by the power cross-spectrum $|v_{ab}(t,f)|^2$, with
\begin{eqnarray}
	\hat{R}_{ab}^2(f, \varphi) &=& \mu_{T_{v_{ab}}(f, \varphi)}^\mathrm{logit}\left[R_{ab}^2(\cdot, f), |v_{ab}(\cdot, f)|^2\right] \label{eqn:R2_fp}
\end{eqnarray}

which is only meaningful in the absence of aliasing where sensors are sufficiently close together, such that the phase domain correctly reflects the direction of propagation.

When numerically calculating the coherence, we find that some values may be equal to 1 or 0, which go to $\pm \infty$ under the logit transformation, contaminating the sum in (\ref{eqn:wavg}) for all other locations in $\Omega$. To circumvent this problem we clip the coherence to a range where the maximum is practically equivalent to 1 and minimum practically equivalent to 0. For the examples presented in this paper we clipped $R^2_{ab}$ to a range of $[0.01, 0.99]$.

\section{Statistics} \label{sec:stats}

For each unit of statistical data, we obtain from the wavelet analysis a 1D curve $\hat{v}(f)$ or $\hat{R}^2(f)$, or a 2D surface $\hat{v}(f, \varphi)$ or $\hat{R}^2(f, \varphi)$. Such a curve or surface is considered to be a response under the influence of a set of predictors which can be any number of categorical or numerical variables specified by the experimental design. We want to measure and compare the effects of the given predictors.

An independent regression model could be fit for each location $x$ in either the frequency $x \in F$ or frequency-phase $x \in F \times H$ domains. However, performing an independent fit at each location would require a multiple-comparison adjustment, and would fail to account for correlations between locations, effectively weakening the power of the analysis.

Instead, we capture correlations between locations by treating the response curves and surfaces as individual functions rather than simply collections of independent points. We model these functions as samples from Gaussian processes, which allow us to specify a formula for correlation between locations, without needing to specify a formula for the shape of the functions themselves.

A Gaussian process (GP) is a probability distribution with an infinite number of random variables, such that any finite set of variables form a multivariate Gaussian distribution. This is achieved by specifying a covariance kernel function $k(x, x')$, which when given a finite set of locations $x \in \{x_1, \dotsc, x_N\}$ allows us to build an $N \times N$ covariance matrix $\Sigma$ with elements $\Sigma_{ij} = k(x_i, x_j)$. We have only finite data, and so the kernel function is evaluated only at the available data locations when fitting the GP. However, we can inspect the GP at any number of arbitrary locations in the kernel's domain, hence the infinite nature of the model as a step beyond a multivariate Gaussian. An analogy is fitting a simple regression line. The line is fit only to a finite set of data, but once we have an intercept $b$ and slope $a$ we can define a function $y(x) = ax + b$, where y-locations can be calculated for any arbitrary set of x-locations, not just those for which we have data. See \cite{rasmussen2006gaussian} for a text book introduction to GPs.

\subsection{Model}

The latent GP function-on-scalar mixed-effect model we use can be written in the form
\begin{eqnarray}
	y_i(x) &\sim& \mathcal{GP}(\eta_i(x), \sigma_i(x, x')) \label{eqn:response} \\
	\sigma_i(x, x') &=& \omega_i(x) \omega_i(x') (k_\sigma(x, x') + \sigma_\epsilon^2) \label{eqn:sigma} \\
	\eta_i(x) &=& \bm{X}_i \bm{\beta}(x) + \bm{Z}_i \bm{b}(x) + o_\eta \label{eqn:eta} \\
	\log(\omega_i(x)) &=& \bm{W}_i \bm{\gamma}(x) + \bm{U}_i \bm{u}(x) \label{eqn:log_omega}
\end{eqnarray}
where $\mathcal{GP}$ represents the Gaussian process distribution, $k_\sigma$ is a kernel function describing the structured $\omega$-standardized noise covariance,  $\sigma_\epsilon^2$ represents unstructured $\omega$-standardized noise variance, and $y_i$ is the response function for observation $i \in \{1, \dotsc, N\}$. The responses are based on the transformed amplitude, $\log(\hat{v})$, or coherence, $\mathrm{\mathrm{logit}}(\hat{R}^2)$. The intuition behind the $\omega$-standardized noise co/variance can be seen by rearranging the terms in (\ref{eqn:response}) and (\ref{eqn:sigma}) to
\begin{eqnarray}
\frac{y_i(x) - \eta_i(x)}{\omega_i(x)} &\sim& \mathcal{GP}(0, k_\sigma(x, x') + \sigma_\epsilon^2) \label{eqn:response_std}
\end{eqnarray}
which facilitates efficient inference by not requiring the structured residuals on the right-hand-side of (\ref{eqn:response_std}) to be sampled, nor requiring a matrix inversion per observation. When evaluating the likelihood specified by (\ref{eqn:response_std}), for the 1D case a simple Cholesky decomposition is sufficient, but for the 2D case an eigendecomposition is needed to separate the kernel functions from the unstructured noise $\sigma^2_\epsilon$ (see \url{https://github.com/lwiklendt/gp_kron_stan} for a Stan model source code example).

In the mean specified by (\ref{eqn:eta}), $\bm{X} \in \mathbb{R}^{N \times P}$ is a design matrix of $P$ population-level predictors (a.k.a. fixed-effects) with $\bm{X}_i \in \mathbb{R}^{1 \times P}$ representing the row vector of predictors pertaining to observation $i$. $\bm{\beta} = (\beta_1, \dotsc, \beta_P)$ is a $P \times 1$ vector of iid latent GPs representing the $P$ population-level effects. $\bm{Z} \in \mathbb{R}^{N \times J}$ is a design matrix of $J$ group-level predictors (a.k.a. random-effects). $\bm{b} = (b_1, \dotsc, b_J)$ is a $J \times 1$ vector of potentially correlated latent GPs representing the group-level effects. Depending on the experimental design, an optional offset term $o_\eta$ is included in (\ref{eqn:eta}) which may be either set to the mean of all $y$ as a way of centring the data, inferred to include a measure of variability in the centring, or given a different value per observation if some measure of exposure needs to be incorporated that would not otherwise fit as its own predictor in $\bm{X}$ or $\bm{Z}$.

Analogous to the predictors $\bm{X}$ and $\bm{Z}$ for the mean, the matrices $\bm{W} \in \mathbb{R}^{N \times Q}$ and $\bm{U} \in \mathbb{R}^{N \times R}$ are respectively the population-level and group-level predictors for the log standard deviation (\ref{eqn:log_omega}), with corresponding effects $\bm{\gamma}$ and $\bm{u}$. An explicit offset term is missing here since such an offset is implicitly handled by the scale of $k_\sigma$.

Each GP function in each vector of population-effects is given an iid prior
\begin{eqnarray}
	\beta_p &\sim& \mathcal{GP}(0, k_{\beta_p}(x, x')) \label{eqn:beta_p} \\
	\gamma_q &\sim& \mathcal{GP}(0, k_{\gamma_q}(x, x')) \label{eqn:gamma_q}
\end{eqnarray}
However, for the vectors of group-effects functions we include correlations between functions via multivariate or multi-output GPs
\begin{eqnarray}
	b_j &\sim& \mathcal{GP}(0, (\Sigma_{\bm{b}})_{j, j'} k_{b_j}(x, x')) \label{eqn:b_j} \\
	u_r &\sim& \mathcal{GP}(0, (\Sigma_{\bm{u}})_{r, r'} k_{u_r}(x, x')) \label{eqn:u_r}
\end{eqnarray}
where $\Sigma_{\bm{b}}$ and $\Sigma_{\bm{u}}$ are covariance matrices dependent on the structure of the $\bm{Z}$ and $\bm{U}$ design matrices. These $\Sigma$ matrices will generally be block-sparse, facilitating efficient computation.

The kernel functions $k_{\{\sigma, \beta, \gamma, b, u\}}(x, x')$ and their parameters, also known as \emph{hyperparameters} of the GPs, will be covered in the next subsection \ref{sec:gpkern}.

The response functions $y_i$, and the design matrices $\bm{X}$, $\bm{Z}$, $\bm{W}$, and $\bm{U}$ are the supplied ``input'' data. The vectors of functions $\bm{\beta}$, $\bm{b}$, $\bm{\gamma}$, $\bm{u}$, and hyperparameters, are to be estimated and correspond to ``outputs'' of the inference. The structure of the design matrices depends on the experimental design, and we find it easiest to derive the design matrices (also known as ``model matrices'') based on formula notation as specified in section 2 of \cite{bates2015fitting}. We provide an example application in section \ref{sec:application} using the formulae (\ref{eqn:formula_mu}) and (\ref{eqn:formula_sigma}).

We are interested in modelling amplitudes or coherences that were calculated using the wavelet transform as described in sections \ref{sec:wt} and \ref{sec:xwt}. To fit amplitudes or coherences over frequencies, $x = f$ is a scalar that represents frequencies. To fit over frequencies and phase-differences, $x = (f, \varphi)$ is a 2D point that represents frequencies in one dimension and phase-differences in the other.

\subsection{Kernel Functions} \label{sec:gpkern}

The form and parameters of the kernel functions $k$ depend on whether the response functions are 1D or 2D. There are many potential kernels to choose from, and they can even be built up from smaller kernels \cite{duvenaud2014automatic}, but for the sake of brevity we will limit our exposition to one concrete kernel function for each type of domain. For the case of 1D curves over frequencies we use a log-space squared-exponential kernel
\begin{eqnarray}
	k(f, f') &=& \tau^2\ \exp \left(- \frac{|\log(f)- \log(f')|^2}{2 \lambda^2} \right) \label{eqn:kern_f}
\end{eqnarray}
with $\lambda$ specifying the lengthscale of the correlation based on the distance $|\log(f) - \log(f')|$ between any two frequencies $f$ and $f'$. At a distance of 0 we have equal frequencies $f = f'$, where the correlation is 1 and covariance is $\tau^2$. As the distance approaches $\infty$ the correlation and covariance approach 0. For the case of 2D surfaces over frequences and phase-differences we use a product of the log-space squared-exponential kernel and a periodic kernel
\begin{eqnarray}
	k(f, \varphi, f', \varphi') = \tau^2\ \exp \left(-\frac{|\log(f)-\log(f')|^2}{2 \lambda^2_f} \right) \nonumber \\ \times \exp \left(- \frac{2 \sin^2\left(\frac{1}{2}|\varphi - \varphi'|\right)}{ \lambda^2_\varphi} \right) \label{eqn:kern_fp}
\end{eqnarray}
where $\lambda_f$ and $\lambda_\varphi$ specify the log-frequency and phase-difference lengthscales. When the difference in phase-differences $\varphi$ and $\varphi'$ is either 0 or $2 \pi$, or any integer multiple of $2 \pi$, then the phase-difference component of the kernel will be 1, identifying the locations $\varphi$ and $\varphi'$.

For equations (\ref{eqn:beta_p}) and (\ref{eqn:gamma_q}), $k$ represents the kernel function used in constructing a \emph{covariance} matrix, but for equations (\ref{eqn:b_j}) and (\ref{eqn:u_r}) $k$ is a kernel function used in constructing a \emph{correlation} matrix by setting $\tau = 1$, since including a free parameter for variance in $k$ would make the model non-identifiable due to the variance parameters already defined in $\Sigma_{\bm{b}}$ and $\Sigma_{\bm{u}}$.

The kernel in (\ref{eqn:kern_fp}) is separable, such that we can write it as
\begin{eqnarray}
	k(f, \varphi, f', \varphi') &=& k(f, f')\ k(\varphi, \varphi) \label{eqn:kern_fp_sep} \\
	k(\varphi, \varphi') &=& \exp \left(- \frac{2 \sin^2\left(\frac{1}{2}|\varphi - \varphi'|\right)}{ \lambda^2_\varphi} \right) \label{eqn:kern_p}
\end{eqnarray}
where with abuse of notation we are identifying kernel functions based on their argument symbols, such that $k(f,f')$ and $k(\varphi, \varphi')$ are different functions, with $k(f,f')$ defined in (\ref{eqn:kern_f}) and $k(\varphi, \varphi')$ defined in (\ref{eqn:kern_p}). Since we can factorise the 2D kernel (\ref{eqn:kern_fp_sep}), we can create a covariance matrix using the Kronecker product of the individual covariance matrices built from kernels (\ref{eqn:kern_f}) and (\ref{eqn:kern_p})
\begin{eqnarray}
	\Sigma &=& \Sigma_F \otimes \Sigma_H \\
	(\Sigma_F)_{ij} &=& k(f_i, f_j)\\
	(\Sigma_H)_{ij} &=& k(\varphi_i, \varphi_j)
\end{eqnarray}

The Kronecker factorisation of the kernel matrices also allows for a substantial speed up in the numerical calculation of the Cholesky and eigen decompositions of the covariance matrices \cite{saatci2012scalable} used in inference.

Prior distributions for hyperparameters $\lambda$ and $\tau$ are experiment dependent, and will in general depend on the scale of the data. For the application presented in section \ref{sec:application}, each $\sigma, \beta, \gamma, b, u$ subscript (omitted for brevity) is treated independently, unless otherwise specified. We used $\lambda \sim \Gamma(2, 1)$ while ensuring $\lambda_\sigma < \lambda_{\{f, \varphi\}}$, for the correlation between $\lambda_f$ and $\lambda_\varphi$ we used $\rho_{f \varphi} \sim \mathrm{Beta}(2, 2)$, and for $\tau$ a half-Student-$t$ distribution with 3 degrees of freedom.

\subsection{Implementation}

A coarse grid was chosen for the functional domain so that posterior sampling could complete within a reasonable time. The grid can be refined relatively quickly after the expensive  sampling step. Rather than the na\"ive linear or cubic interpolation, we can use GP prediction such that the covariance between locations is faithfully preserved in the refinement.

Given a vector of $N$ grid coordinates $\mathbf{x}$, a vector of $M$ refined coordinates $\mathbf{x}_*$, a vector of $N$ function values $\mathbf{y}$ corresponding to $\mathbf{x}$, and the kernel function $k$, then we can produce a vector of $M$ refined function values $\mathbf{y}_*$ at $\mathbf{x}_*$ with
\begin{eqnarray}
	\mathbf{y}_* &=& \Sigma(\mathbf{x}_*, \mathbf{x}) \Sigma(\mathbf{x}, \mathbf{x})^{-1} \mathbf{y} \label{eqn:refine}
\end{eqnarray}
where $\Sigma(\mathbf{x}, \mathbf{x})$ is the $N \times N$ covariance matrix obtained by applying $k$ to the coordinates in $\mathbf{x}$, and $\Sigma(\mathbf{x}_*, \mathbf{x})$ is the $M \times N$ matrix given by the covariances obtained by applying $k$ to $\mathbf{x}_*$ and $\mathbf{x}$.

Note, for the 2D case we can take advantage of
\begin{eqnarray}
	(\Sigma_F \otimes \Sigma_H)^{-1} &=& \Sigma_F^{-1} \otimes \Sigma_H^{-1}
\end{eqnarray}

We use the Hamiltonian Monte Carlo sampler from the Stan package to obtain a posterior distribution of GPs which can be inspected to detect where and how locations may differ between various categorical predictors such as anatomical regions, treatment periods, patient types, and between various numerical predictors such as age, weight, caloric content of a meal, etc.

\section{Application} \label{sec:application}

We applied the method to colonic manometry data recorded from the descending and sigmoid colon of 11 healthy volunteers during 1hr preprandial and postprandial periods. Data from these studies have been published previously \cite{dinning2014quantification}. Using formula notation
\begin{eqnarray}
	\eta &\sim & \mathtt{reg * meal + (reg + meal\ |\ subj)} \label{eqn:formula_mu} \\
	\log(\omega) &\sim & \mathtt{reg * meal + nchan} \label{eqn:formula_sigma}
\end{eqnarray}
the design matrices $\bm{X}$ and $\bm{Z}$ are constructed from formula (\ref{eqn:formula_mu}), and $\bm{W}$ and $\bm{U}$ from (\ref{eqn:formula_sigma}), according to the construction process described in \cite{bates2015fitting}, where the $\log$ in (\ref{eqn:formula_sigma}) is a transformation of $\omega$. The $\mathtt{reg}$ predictor is a categorical variable with two levels, one for each of the recorded regions of the colon: descending and sigmoid. The $\mathtt{meal}$ predictor is a binary variable indicating whether a recording was obtained during the preprandial ``0'' or postprandial ``1'' state, corresponding to a meal effect. The categorical variable $\mathtt{subj}$ identifies the individual subject of a recording.

The $\mathtt{nchan}$ predictor in (\ref{eqn:formula_sigma}) is a real-valued standardised count of the number of sensors (or channels) in the recording, which varies per subject per region. When computing weighted-averages over time as specified in sections \ref{sec:wt} and \ref{sec:xwt}, we average not only over time but over both time and channels by effectively flattening the wavelet results into a single channel of length $c |T|$, where $c$ is the number of channels and $|T|$ is the number of time samples. Fewer channels are expected to result in a greater variation in the global averages, which is why we included it as a confounding factor of the signal variance. We set $\bm{U} = 0$ with the formula (\ref{eqn:formula_sigma}) since we don't have repeated measurements, and so a within-subject variation is poorly identified.

\begin{figure*}
	\begin{subfigure}[t]{\columnwidth}
		\centering
		\includegraphics[width=\textwidth]{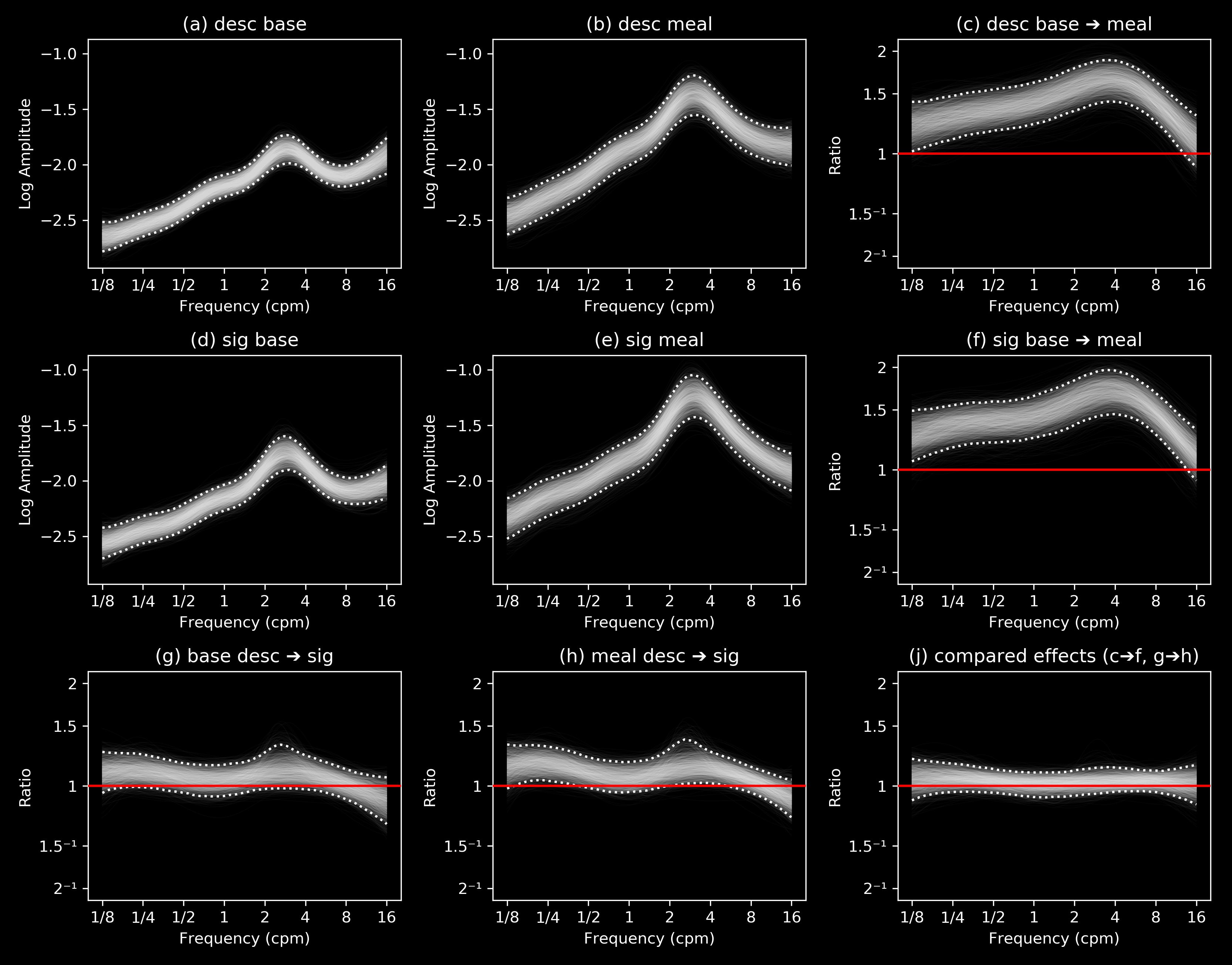}
		\caption{Amplitude over frequency}
		\label{fig:amp1d}
	\end{subfigure}
	~
	\begin{subfigure}[t]{\columnwidth}
		\centering
		\includegraphics[width=\textwidth]{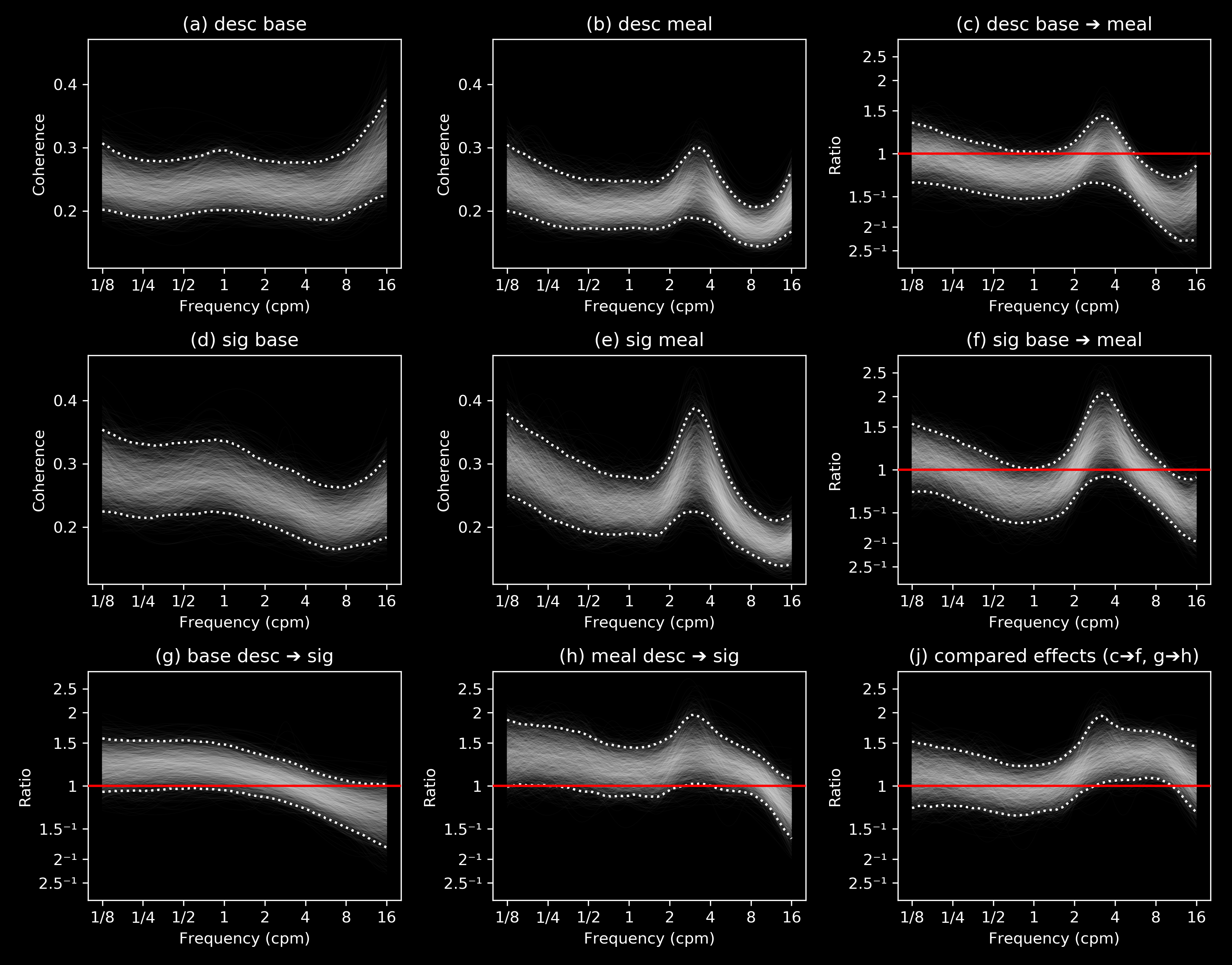}
		\caption{Coherence over frequency \vspace{2mm}}	
		\label{fig:coh1d}
	\end{subfigure}

	%\iflinenumbers \internallinenumbers \fi
	
	\caption{Various ways of inspecting the latent $\bm{\beta}$ functions of equation (\ref{eqn:eta}). In each panel, cell (a) corresponds to the intercept, (c) is the meal effect, (g) is the sigmoid effect, and (j) is the sigmoid meal interaction effect. All other cells are additive combinations of those four latent $\bm{\beta}$ functions. Due to the log and logit transforms of the responses, the effects are logarithmically additive, which are represented as ratios. We plot all 2000 posterior samples as overlaid thin white lines, with 95\%  ``credible intervals'' outlined by white dotted lines. A substantial meal effect is evident in the amplitude responses (i) where the posterior curves in cells (c) and (f) are clearly above a ratio of 1, with greatest effect around the 3-4 cycles-per-minute (cpm) frequency. In (ii), although there is no such clear meal effect, we can see that in the sigmoid region the meal effect (f) around 3-4 cpm is increasing compared to the decreasing effect adjacent at around 1cpm and above 8cpm. We also notice the difference in meal effect between the descending and sigmoid regions (j) as an increase in coherence between 3-10cpm. An uptick near 16cpm in (a, b, d, e) is likely due to respiration artefact.}
	
	\label{fig:results1d}
\end{figure*}

\begin{figure*}
	\begin{subfigure}[t]{\columnwidth}
		\centering
		\includegraphics[width=\textwidth]{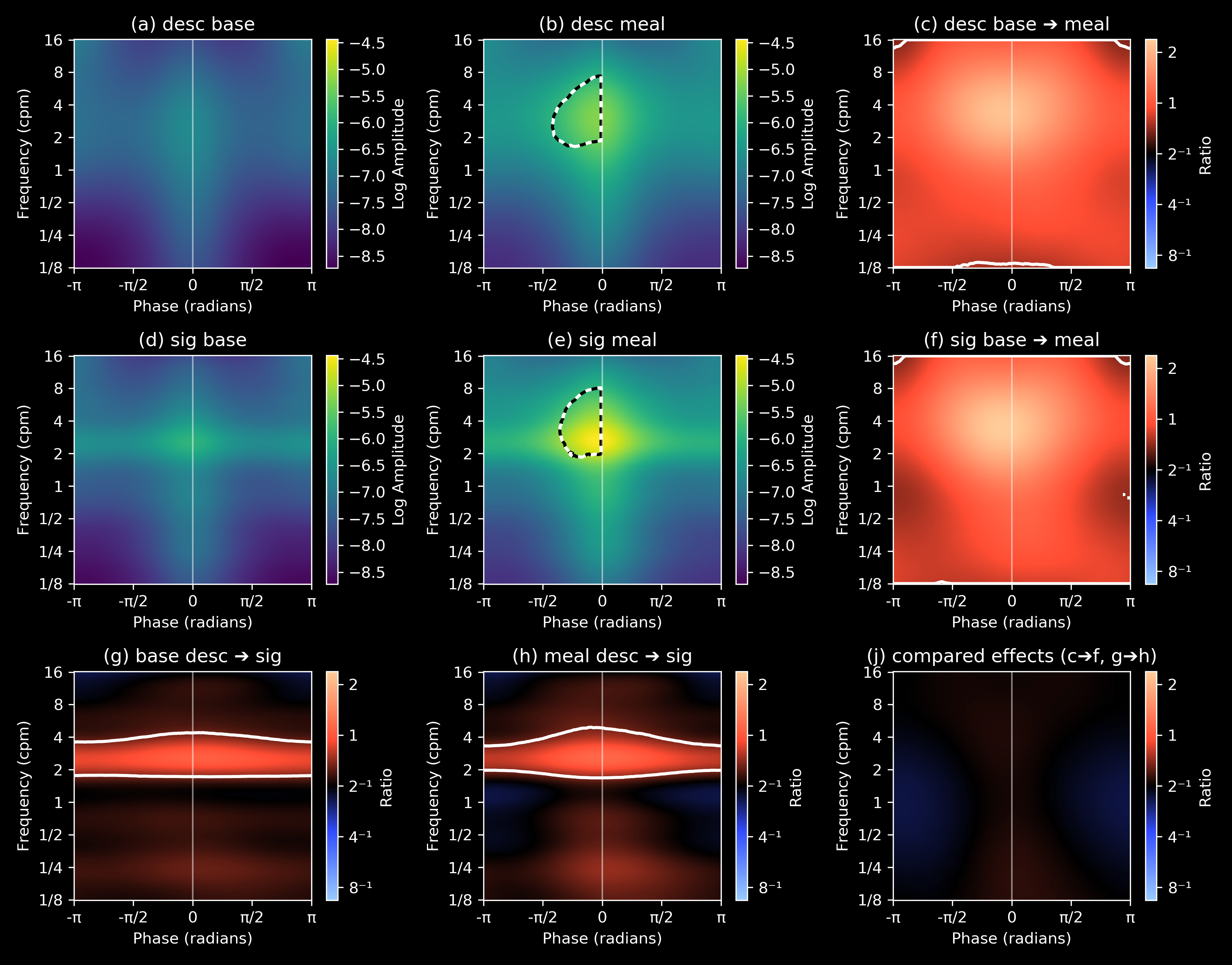}
		\caption{Amplitude over frequency and phase}
		\label{fig:amp2d}
	\end{subfigure}
	~
	\begin{subfigure}[t]{\columnwidth}
		\centering
		\includegraphics[width=\textwidth]{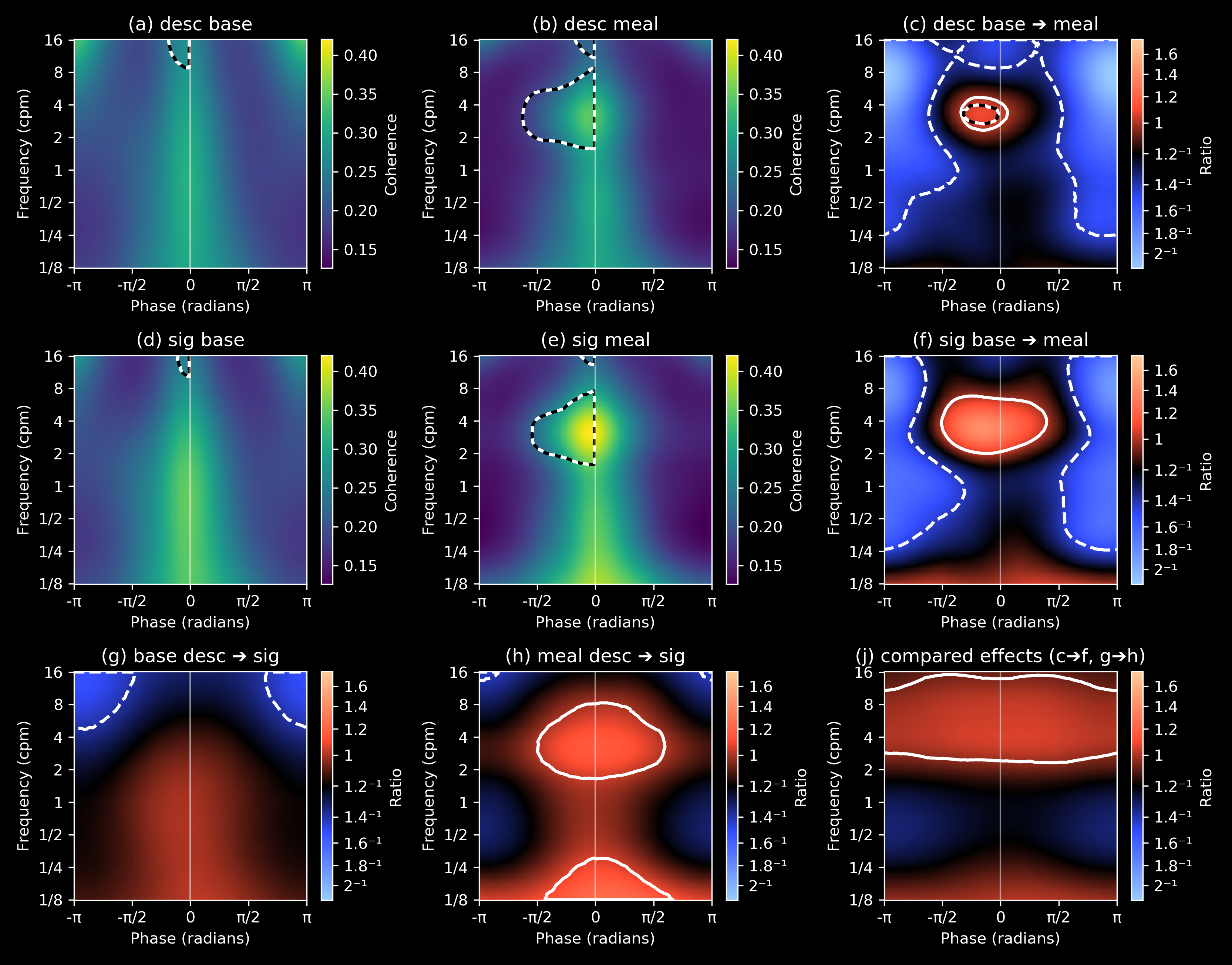}
		\caption{Coherence over frequency and phase}
		\label{fig:coh2d}
	\end{subfigure}

	%\iflinenumbers \internallinenumbers \fi

	\caption{Panels are arranged in the same way as described in Fig. \ref{fig:results1d}. The 2D nature of the responses makes it impossible to legibly plot all 2000 posterior samples overlaid, and we instead plot the per-location median of the samples. Regions encircled in solid white lines have their 95\% credible intervals above a ratio of 1, and regions encircled in dashed white lines have their 95\% credible intervals below a ratio of 1. Regions encircled by black-and-white dashed lines correspond to regions where 97.5\% of the samples are above the samples at the same frequency but opposite phase, and can be interpreted as predominantly propagating activity.}
	
	\label{fig:results2d}
\end{figure*}

Four types of responses were analysed, given by the 1D and 2D amplitudes from equations (\ref{eqn:gws}) and (\ref{eqn:v_fp}), and the 1D and 2D coherences from equations (\ref{eqn:R2f}) and (\ref{eqn:R2_fp}). The amplitudes were log-transformed and coherences logit-transformed to obtain the $y$'s in model (\ref{eqn:response}-\ref{eqn:log_omega}). For the 1D responses 29 frequency-bins were used, and for the 2D responses 15 frequency-bins and 16 phase-bins were used. After sampling from the posterior, 200 frequency bins and 201 phase bins were used in refinement via GP interpolation (\ref{eqn:refine}), with results plotted in Fig. \ref{fig:results1d} for 1D responses and Fig. \ref{fig:results2d} for 2D responses.

For each response type, the Hamiltonian Monte Carlo run consisted of 200 warm-up iterations and 500 sampling iterations over 4 randomly-initialised chains resulting in 2000 samples from the posterior distribution. We used an adapt-delta of 0.95. Diagnostics showed no divergent transitions, a top tree depth of 9, and visual inspection of trace plots showed good convergence that was validated by an $\hat{R} \approx 1$\footnote{Here $\hat{R}$ refers to the convergence diagnostic in Stan, and not our coherence measure from equations (\ref{eqn:R2f}) and (\ref{eqn:R2_fp}).}. Data appeared consistent with the posterior predictive distribution. On a 2013 MacBook Pro with 2.7GHz Core i7 running macOS Mojave with 16GB 1600MHz RAM using PyStan\footnote{Stan Development Team. 2018. PyStan: the Python interface to Stan, Version 2.18. http://mc-stan.org} with 4 parallel CPU cores (1 per chain), each of the 1D response types completed sampling in $\approx 15$ minutes, and each of the 2D response types completed in $\approx 3$ hours. The computation process from raw pressure recording to time-averaged wavelet spectra (Fig. \ref{fig:wavelet}) for each of the 42 individual observations took $\approx 30$ seconds.

\section{Discussion} \label{sec:discussion}

We have presented a method for analysing spatiotemporal manometry data by computing various time-averaged spectra and using them as responses in a functional mixed-effects model, inferred via HMC. This has allowed us to identify and contrast features in the spectra between various predictors of each spectrum.

Although the statistical model presented was developed to analyse quasiperiodic spatiotemporal manometry data, the model can be used (with appropriate choice of kernel function) to practically any kind of functional data that can be represented on a shared grid.

For the statistical analysis, we are not limited to testing for differences across predictors within the same locations. An example of this is already shown in Fig. \ref{fig:results2d} where the black-and-white hatched lines encircle regions where differences are inspected not between predictors but between phase locations within the same set of predictors. Since HMC produces samples from a distribution over all parameters $\bm{\theta}$ given the data $D$, that is $p(\bm{\theta} | D)$, we are free to interrogate this distribution by applying  any well-posed\footnote{Well-posed questions are in the form of an expectation over a function $f$ applied to the random variable $\bm{\theta}$, with distribution $p(\bm{\theta} | D)$. This is achieved by summing $f$ over the $n$ samples $\{\bm{\theta}_1, \bm{\theta}_2, \ldots, \bm{\theta}_n\}$ from the posterior distribution, formally, $\mathbb{E}[f(\bm{\theta})] \approx \frac{1}{n}\sum_{i=1}^n f(\bm{\theta}_i)$. This is not as strong of a limitation as may first appear. For example, $f$ could include the $\mathrm{argmax}$ function over a subset of parameters, allowing us to obtain the peak amplitude or coherence as a distribution over frequencies, and to potentially test how this peak frequency depends on predictors.} question of interest to the samples.

The method presented in \cite{wiklendt2013classification} was a first step in automated analysis of colonic propagation, but reduced the analysis to a single indicator that could discern neither speed of propagation nor frequency of activity. The results herein are confirmed by our previous manual analysis in \cite{dinning2014quantification}. In that publication we demonstrated significant increased in the post prandial colonic cyclic motor pattern. That motor pattern consisted primarily of propagating pressure waves at a 3-4cpm frequency that travelled mostly in a retrograde direction. In our previous publication we also showed that the majority (76\%) of the cyclic motor patterns was identified in the sigmoid colon. With our novel technique, in this paper these physiological features are all shown. The manual analysis in our previous paper \cite{dinning2014quantification} took several weeks to perform. To obtain similar results with the developed analysis in this paper took under 7 hours. Thus the detailed analysis provided by our developed technique is orders of magnitude beyond the methods currently available in both detail, speed of analysis, and manual labor saved. A postprandial increase in colonic activity was also shown in \cite{dinning2015colonic} using FFT and \cite{dinning2016high} using wavelets but without the rigorous statistical analysis the current paper provides. 

%There currently exists no exposition of a detailed automated method for analysing cyclic propagating colonic manometry data. The method presented in \cite{wiklendt2013classification} was a first step in automated analysis of colonic propagation, but reduced the analysis to a single indicator that could discern neither speed of propagation nor frequency of activity. The results herein are validated with the manual analysis of \cite{dinning2014quantification} that also showed an increased postprandial colonic activity, and that of \cite{dinning2015colonic} using FFT and \cite{dinning2016high} using wavelets which suggest a postprandial increase at around the 3-4cpm frequency of activity, but without the rigorous statistical analysis the current paper provides. With our novel technique we can now show a significant increase in the coherence of 2-6cpm activity, that is also significantly greater in retrograde than the anterograde direction in the descending colon. This analysis corroborates with hypotheses of a ``break'' mechanism in the distal colon which prevents content from prematurely escaping \cite{rao1996periodic}. Such detailed analysis is orders of magnitude beyond the methods currently available in both detail, speed of analysis, and manual labor saved.

\subsection*{Caveats and Limitations}

When applying synchrosqueezing, we sometimes notice frequency-edge artefacts for activity close the the highest or lowest frequencies in $\mathcal{F}$. This may be due to an insufficient source of instantaneous frequency to be redistributed near the edge frequencies with equation (\ref{eqn:syncsqz}). We suggest to choose sufficiently high and low scales to capture activity at least one octave above and below the highest and lowest frequencies to be examined, as long as the data is of sufficient sampling frequency and duration to permit such a margin.

In the application presented here, our data was recorded from a catheter with equally-spaced sensors. For unequally-spaced sensors, sensor pairs at different separations showing the same phase-difference will not represent the same propagation speed. Either a per-sensor-pair adjustment to the phase-difference should be made, or the inverse-velocity (pace) should be considered as a substitute for the phase-difference. This will require choosing a more appropriate phase-domain kernel since the phase-dependent result will no longer neatly fall into a periodic kernel with a constant period over all frequencies.

There is a clear use of amplitude-weighted coherence, equations (\ref{eqn:R2f}) and (\ref{eqn:wavg}), for cases where sensors are too far apart to obtain a meaningful measure of phase-difference. However, the amplitude-weighted coherence is also useful even when sensors are sufficiently close, if the experiment requires periods of quiescence to be ignored. We may want to focus only on periods of higher activity (higher amplitudes) if quiescent periods (lower amplitudes) occur at random periods in the signal and/or have coherences that are unrepresentative or independent of any predictors.

If the amplitude-weighted coherence, (\ref{eqn:R2f}) and (\ref{eqn:R2_fp}), is applied to a weak signal where a momentary but large amplitude is present, then that momentary event will dominate the rest of the recording. To account for the fact that substantially fewer time samples influence the calculation of the coherence in such cases, a predictor for the signal variance (\ref{eqn:formula_sigma}) can be added, which quantifies the proportion of the signal from which coherence is effectively retained due to amplitude-weighting.

Formula (\ref{eqn:response_std}) allows for GP residuals. It is unclear how one could extend the responses to other distributions, such as Poisson or Bernoulli, while retaining residual correlation. Perhaps one would need to either sample latent structured residuals, or abandon the residual correlation altogether.

Data where some grid points have missing values is not handled. Imputing the missing values could be a solution, achieved by sampling from the residual GP at the missing locations. This would require computing the mean and covariance of the residual GP at the missing values per observation (Algorithm 2.1 in \cite{rasmussen2006gaussian}). Had we not used structured residuals, such imputation would be trivial to implement and computationally inexpensive.

%%%%    NOTE: use the following to generate a *.bbl    %%%%

\bibliographystyle{IEEEtran_keep_case}

%\bibliography{/Users/luke/Dropbox/Literature/lukasz}
%\bibliography{D:/Dropbox/Literature/lukasz}

%%%%    copy the *.bbl to same locations as this tex file for submission    %%%%

\input{wavelet.bbl}

%%%%    comment-out one of the above two blocks to decide bib source    %%%%

\end{document}

%% file: wavelet.bbl
% Generated by IEEEtran.bst, version: 1.12 (2007/01/11)